\documentclass[twocolumn]{revtex4-2}

\usepackage{graphicx}
\usepackage{dcolumn}
\usepackage{bm}
\usepackage{amssymb}

\usepackage{amsmath}

\AtBeginDocument{%
  \setlength{\abovedisplayskip}{6pt}
  \setlength{\belowdisplayskip}{6pt}
  \setlength{\abovedisplayshortskip}{6pt}
  \setlength{\belowdisplayshortskip}{6pt}
}

\begin{document}

\title{\textbf{Covariance and the use of the Schr\"odinger equation in quantum field theory} }

\author{J.D. Franson}
 \affiliation{Physics Department, University of Maryland Baltimore County, Baltimore, MD USA}

\date{\today}

\begin{abstract}

The Schr\"odinger equation is not covariant. Nevertheless, quantum field theory is often formulated using the Schr\"odinger equation to describe the time evolution of the system, which is equivalent to using Feynman path integrals. It is well known that scattering theory gives covariant results provided that the interaction vanishes in  the asymptotic limit of $t \rightarrow \pm \infty$, but there are other situations of interest that do not satisfy those conditions.  As an example, the  Aharonov-Bohm effect is derived here using second-quantized field theory to describe all the electrons as well as the electromagnetic field, which allows the effects of retarded vector potentials to be calculated in a straightforward way using the Feynman propagator.  The results are in agreement with the usual expression for the Aharonov-Bohm effect when the retardation of the electromagnetic field is negligible, but they predict a fractional phase shift, a lack of covariance, and violations of causality when retardation effects are significant. These results suggest that quantum field theory based on the Schr\"odinger equation or Feynman path integrals is incomplete in the sense that it cannot give a correct description of all observable phenomena.

\end{abstract}

\maketitle


\section{INTRODUCTION}

The Schr\"odinger equation is not covariant due to the special role of the time $t.$  Nevertheless, quantum field theory is often formulated using the Hamiltonian and the Schr\"odinger equation to describe the time evolution of the system \cite{weinberg1995,peskin1995}, which is equivalent to using the Lagrangian and Feynman path integrals \cite{feynman1948,Feynman1965}. For example, the Dyson formula for the time evolution operator $\hat{U} (t,t')$ was derived using the interaction Hamiltonian and the Schr\"odinger equation \cite{dyson1949}. It is well known \cite{Dirac1927,dyson1949,lehmann1955,bogoliubov1959,haag1958quantum,schweber1962} that the use of the Schr\"odinger equation gives covariant results for scattering theory provided that the interaction vanishes in the initial and final states in the limit of $t \rightarrow \pm \infty,$ as can be seen using the equivalent path integral formulation. 

This paper considers a different situation in which the interaction is zero at all locations in a coordinate frame $O$ at the time $t_m$ of a measurement, while the interaction is nonzero in some regions of a different coordinate frame $O'$ at the corresponding time $t_m'.$   This is possible due to the relative nature of simultaneity inherent in a Lorentz transformation.  The fact that the interaction is zero in coordinate frame $O$ avoids the usual concern that the operator representing the measurement may not be well defined in the presence of an interaction, while the fact that the interaction is nonzero in coordinate frame $O'$ invalidates the usual proof of covariance as shown in more detail below.

As an example, the magnetic Aharonov-Bohm effect \cite{Aharonov1959,aharonov1961,baym1973,boyer1973,webb1985,Tonomura1986,Peshkin1989,Kurizki1991,bergman1994,bachtold1999,gomes2000,vaidman2012role,Pearle2017a,Saldanha2023} is derived using second-quantized field theory to describe the electromagnetic field and all the electrons in a consistent way.  The Aharonov-Bohm effect is of fundamental interest because it shows that the vector potential can have an observable effect on an electron that is traveling through an interferometer that encloses a solenoid, even though the electric and magnetic fields are zero along the path of the electron as illustrated in Fig. 1.  This suggests that the vector potential may be more fundamental than the electric and magnetic fields.

\begin{figure}[tbp]
\centering
\includegraphics[width=0.28\textwidth]{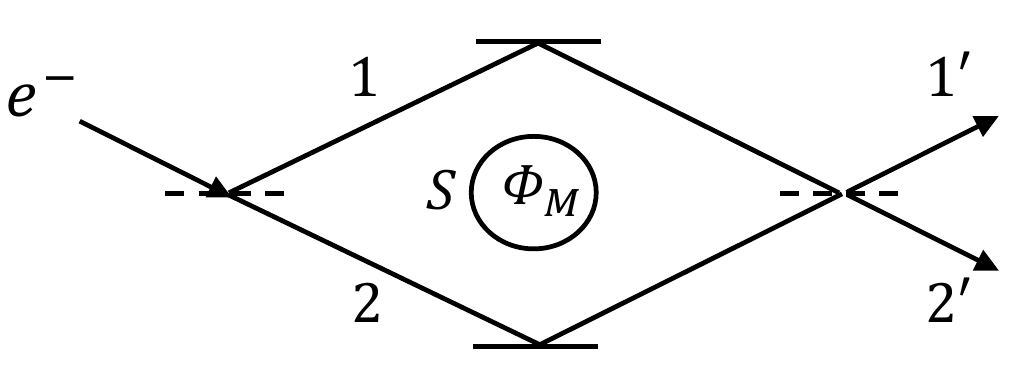}
\qquad
\caption{Magnetic Aharonov-Bohm effect.  The field of an incident electron $(e^-)$ is split into two paths 1 and 2 by a beam splitter (dashed line).  The vector potential from an enclosed solenoid $S$ with magnetic flux $\Phi_M$ can change the relative phase  between the two paths, even though the electric and magnetic fields are zero outside of the solenoid.}
\label{Fig 1}
\end{figure}

The Aharonov-Bohm phase shift is derived here using  perturbation theory and the properties of the Feynman propagator for the electromgnetic field. The phase shift in the interferometer can be extracted directly from the time-evolution operator $\hat{U} (t,t')$, which is the most straightforward approach.  The same results are also obtained using in-in (Schwinger-Keldysh) perturbation theory \cite{schwinger1961,Keldysh1965,baym1961,baym1962,Kadanoff1962,rammer2007}, which is more rigorous.

The phase shift $\phi_i$ along a given path $i$ ($i=1,2$) through the interferometer is shown to have the form
 \begin{eqnarray}
   \phi_i =  \frac{1}{2}\ \int d^3 \mathbf{r} \, dt \, \left[ 
      \boldsymbol{J}_e   \cdot \boldsymbol{\mathcal{A}}_{S} +   \boldsymbol{J_S}   \cdot \boldsymbol{\mathcal{A}}_{e} \right] .
     \label{result12p}
 \end{eqnarray}
Here ${\boldsymbol{J}}_e(\mathbf{r},t)$ is the expectation value of the current produced by an electron in path $i$ through the interferometer while $\boldsymbol{J}_S(\mathbf{r},t)$ is the expectation value of the current in the solenoid.
 $ \boldsymbol{\mathcal{A}}_e (\mathbf{r},t) $ and $ \boldsymbol{\mathcal{A}}_S (\mathbf{r},t)$ are the expectation values of the retarded vector potentials produced by the electron in the interferometer and those in the solenoid, respectively. (Natural units with $c=\hbar=1$ will be used.) It can be shown that the contributions from the two terms on the right-hand side of Eq.\ \eqref{result12p} are equal for a quasistatic field where retardation effects are negligible, in which case the phase shift reduces to the usual expression for the Aharonov-Bohm effect. 

 The factor of $\frac{1}{2}$ and the symmetric form of Eq.\  \eqref{result12p} follow directly from the properties of the Feynman propagator as shown below. The first term in Eq.\  \eqref{result12p} corresponds to the Feynman diagram in Fig. 2(a) while the second term corresponds to the diagram in Fig. 2(b).

\begin{figure}[tbp]
\centering
\includegraphics[width=0.39\textwidth]{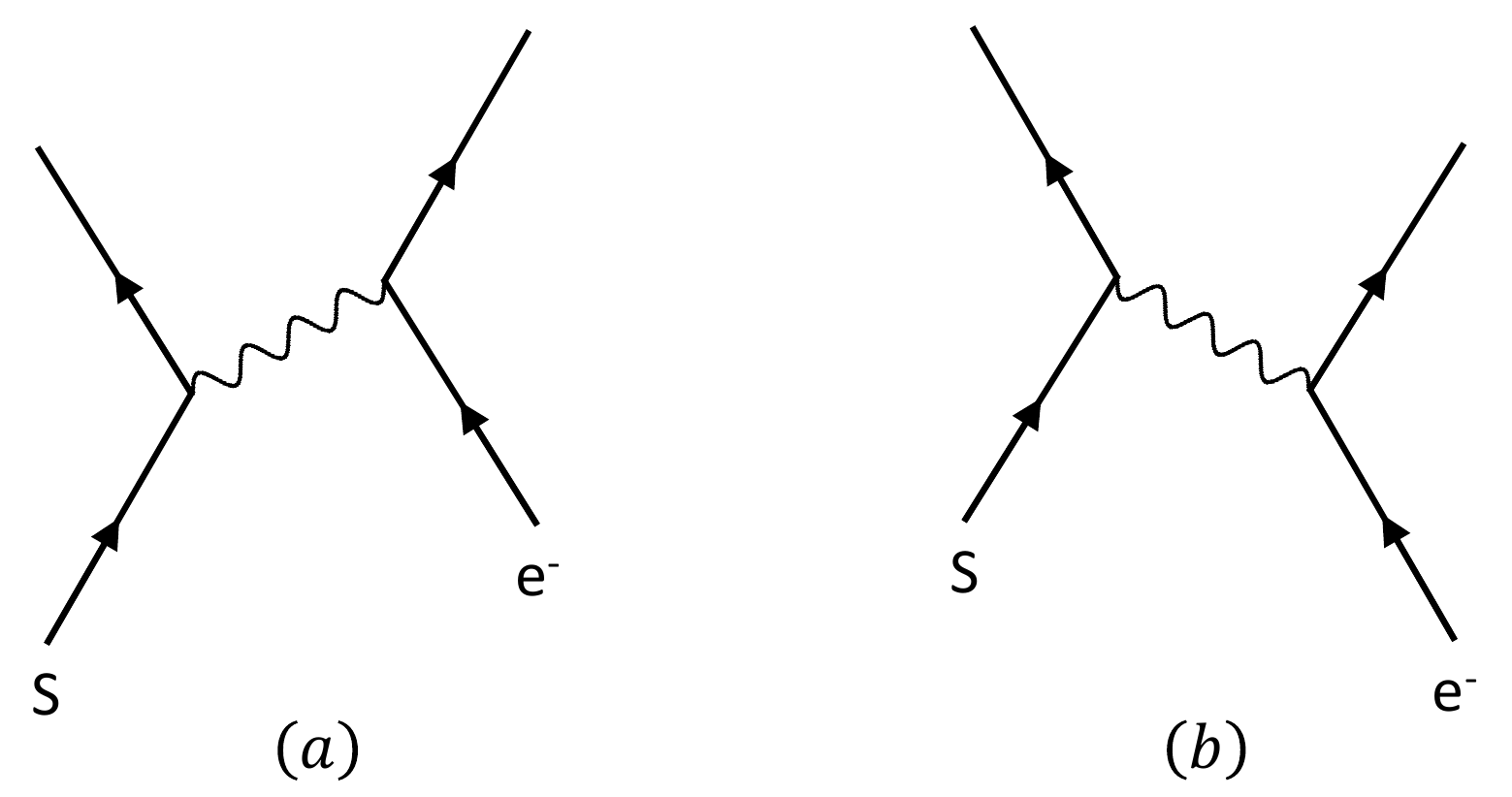}
\qquad
\caption{Feynman diagrams contributing to Eq.\ \eqref{result12p}. (a) Creation of a photon by an electron in the source S followed by its annihilation by the electron in the interferometer, which gives the first term in Eq.\ \eqref{result12p}. (b) Creation of a photon by the electron in the interferometer followed by its annihilation by an electron in the source, which gives the second term in Eq.\ \eqref{result12p}.  
\label{Fig 2}}
\end{figure}

 The symmetric form of Eq.\  \eqref{result12p} can  be  intuitively understood by considering the use of the Feynman path integral approach, where the probability amplitude for each path through the configuration space of the system is given by $e^{i \mathcal{S}}$.  The action $\mathcal{S}$ is that of the entire system, and the two terms in Eq.\ \eqref{result12p} correspond to the action associated with the electron in the interferometer and those in the solenoid. As a result, the output of the interferometer depends on the total phase shift of the system for an electron traveling along path $1$ or $2$.  This dependence on the overall phase of the system has been demonstrated previously in nonlocal interferometer experiments that violate Bell's inequality \cite{franson1989,Gisin1998,ribordy2000,brougham2013,bulla2023,fallon2025}. 

The results of the calculation are more problematic when retardation effects are significant.  In that case, the output of the interferometer can depend on changes in the value of $\boldsymbol{J}_S$ in Eq.\ \eqref{result12p} even when they are spacelike separated, and conventional field theory predicts a fractional Aharonov-Bohm effect, a lack of covariance, and violations of causality.  

These results are in disagreement with the generally-accepted proof of causality based on the
commutation or anticommutation relations of the field operators.  This discrepancy can be understood by considering the role of the commutation relations in the interaction picture using in-in perturbation theory \cite{schwinger1961,Keldysh1965,baym1961,baym1962,Kadanoff1962,rammer2007}.  The output of the interferometer is determined by the phase shift in the field $\hat{\psi} (\mathbf{r}_i,t)$ in each path $i$, whose expectation value is given by
\begin{eqnarray} 
\langle \Phi(t) | \hat{\psi} (\mathbf{r}_i,t) | \Phi(t) \rangle &=& \langle \Phi_0 | \hat{U}^\dagger \hat{\psi} (\mathbf{r}_i,t) \hat{U} | \Phi_0 \rangle 
 .  
    \label{p1prime38tt}
\end{eqnarray} 
Here $| \Phi_{0} \rangle$ is the initial state of the system in the interaction picture.

To lowest order, Eq.\ \eqref{p1prime38tt} can be used to show that the change in the phase of the field due to the interaction is given by
\begin{eqnarray} 
\delta \langle \hat{\psi} (\mathbf{r}_i,t) \rangle 
  &=&  \langle  \Phi_0 |  \, [ \hat{\psi} (\mathbf{r}_i,t), \hat{{U}}^{(2)}   ] \, |  \Phi_{0} \rangle  .  
    \label{p1prime459}
\end{eqnarray}
Here $\hat{{U}}^{(2)} $ is the second-order contribution to the time-evolution operator in the interaction picture, which is given by
\begin{eqnarray}
     \hat{{U}}^{(2)} & = & \frac{i}{2}\  \int d^3 \mathbf{r}' \,\int_{t_i}^{t} dt' \int d^3 \mathbf{r}''  \int_{t_i}^{t'} dt''   \nonumber \\[2pt]
     & \times &  \hat{j}_l (\mathbf{r}',t')  \, G_R^{lm} (\mathbf{r'},t';  \mathbf{r}'',t'')   \hat{j}_m (\mathbf{r}'',t''). 
     \label{result77}
 \end{eqnarray}
 $G_R^{lm} (\mathbf{r}',t';  \mathbf{r}'',t'')$ is the retarded propagator for the electromagnetic field, where the expectation value in the initial (vacuum) state of the electromagnetic field has already been taken.  $\hat{j}_l (\mathbf{r}',t')$ and $\hat{j}_m (\mathbf{r}'',t'')$ correspond to the current operator evaluated at two different spacetime points.  

  As shown below, the commutator of $\hat{\psi} (\mathbf{r}_i,t)$ with $\hat{j}_l (\mathbf{r}',t')$ in Eq.\ \eqref{result77} corresponds to the process shown in Fig. 2(a), which gives the first term in Eq.\ \eqref{result12p}.  The commutator of $\hat{\psi} (\mathbf{r}_i,t)$ with $\hat{j}_m (\mathbf{r}'',t'')$ corresponds to the process shown in Fig. 2(b), which gives the second term in Eq.\ \eqref{result12p}. These commutators are nonzero when either $\mathbf{r}'$ or $\mathbf{r}''$ is located in path $i$ through the interferometer, in which case they are timelike separated from $\hat{\psi} (\mathbf{r}_i,t)$.  

The violation of causality can now be understood from the fact that the field in the interferometer is determined by its commutator with the time-evolution operator, not its anticommutator with the field in the source, as shown in Eq.\ \eqref{p1prime459}.  The commutator with $\hat{{U}}^{(2)}$ can be nonzero even when the interferometer and the current in the source are spacelike separated with $\{ \hat{\psi} (\mathbf{r}_i,t), \bar{\psi}(\mathbf{r}_S,t') \}=0$.

These results suggest that quantum field theory based on the Schr\"odinger equation or Feynman path integrals is incomplete \cite{einstein1935} in the sense that it cannot give a correct description of all observable phenomena.

The remainder of this paper is organized as follows. Covariant perturbation theory for the usual case where the interaction vanishes for $t \rightarrow \pm \infty$ is reviewed in Section II. The Aharonov-Bohm phase shift is derived in Section III using perturbation theory and the properties of the Feynman propagator.  Section IV considers the case of quasistatic fields and the role of the overall phase of the quantum state of the system.  The results of the calculations are further discussed in Section V, including the prediction of a fractional Aharonov-Bohm effect for retarded vector potentials and violations of causality.  The role of the commutation and anticommutation relations in causality are discussed in more detail in Section VI, while the results are summarized in Section VII.  The potential effects of entanglement between the electrons are discussed in Appendix A.  The Aharonov-Bohm effect is rederived using in-in (Schwinger-Keldysh) perturbation theory \cite{schwinger1961,Keldysh1965,baym1961,baym1962,Kadanoff1962,rammer2007} in Appendix B, which gives the same results as are obtained in the text.  

\section{COVARIANT PERTURBATION THEORY}

The need for a covariant form of perturbation theory has been recognized since the early days of quantum field theory \cite{Dirac1927,tomonaga1946,schwinger1948,lehmann1955,bogoliubov1959,haag1958quantum,schweber1962}.  The potential problem in using the Schr\"odinger equation can be seen in the spacetime diagram of Fig.\ 3(a), where a measurement is performed at time $t_m$ in coordinate frame $O$ and at the corresponding time $t_m '$ in a different coordinate frame $O'.$  The Schr\"odinger equation is integrated up to time $t_m$ in $O$ and up to time $t_m '$ in $O'$, as represented by the red (solid) and blue (dashed) lines.  

The interaction is nonzero at the time of the measurement in this example, as represented by the shaded circle.  As a result, the interaction in region A is included in the integral over time in $O$ but not in $O'$, while the interaction in region B is included in $O'$ but not in $O$.  Thus the quantum states in $O$ and $O'$ are fundamentally different and they cannot be related by a simple Lorentz transformation in general.  In addition, the operator that represents the measurement in the Heisenberg or interaction pictures is not given by its simple form in the Schr\"odinger picture.  

Both of these difficulties can be avoided if the interaction vanishes at the time of the measurement as illustrated in Fig. 3(b), as is the case in scattering theory.  The integral over time includes the same interactions in both coordinate frames, so that the use of perturbation theory is covariant provided that the interaction Hamiltonian  $H'$ is a Lorentz scalar.   This can be shown directly using the Hamiltonian and the Schr\"odinger equation \cite{schweber1962}.  

The covariance of this approach is more apparent using Feynman path integrals, where the probability amplitude associated with each path through the configuration space of the system is  $e^{i\mathcal{S} }$.  The action $\mathcal{S}$ is given as usual by
\begin{equation}
 \mathcal{S}=\int  d^4 x  \, \mathcal{L} ,
 \label{action}
\end{equation}
where  $\mathcal{L}$ is the Lagrangian density.  The action will be a Lorentz invariant if the interaction vanishes for  $t_0 \rightarrow -\infty$  and  $t_f \rightarrow +\infty$ , provided that the Lagrangian density is also a Lorentz invariant. In that case, each path will be weighted the same in all coordinate frames which ensures that the scattering amplitudes will be covariant.

\begin{figure}[tbp]
\centering
\includegraphics[width=0.47\textwidth]{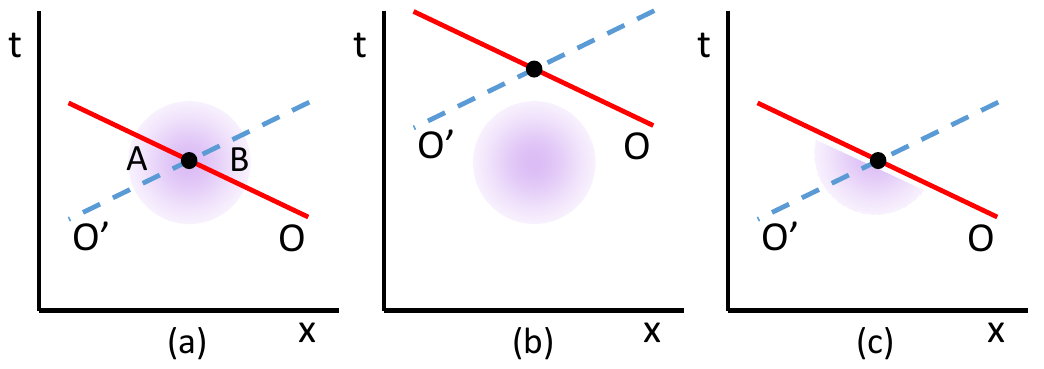}
\qquad
\caption{Covariance and the use of the Schr\"odinger equation.  (a)  A measurement is made at time $t_m$ in an inertial coordinate frame $O$ and at the corresponding time $t_m '$ in a second coordinate frame $O'$ moving with a velocity $v$ relative to $O$, as indicated by the black dot.  The presence of an interaction is represented by the shaded circle.  The Schr\"odinger equation is integrated over all space up to time $t_m$ in $O$ as indicated by the red (solid) line, while it is integrated up to time $t_m '$ in $O'$ as indicated by the blue (dashed) line.  (The coordinates $x$ and $t$ in the figure are those observed in a third coordinate frame with an intermediate velocity of $v/2$.)  The interaction in region A is present in the integral in $O$ but not in $O'$, while the interaction in region B is present in the integral in $O'$ but not in $O$. This results in a lack of covariance.  (b)  The  measurement occurs at a later time so that the entire interaction is included in both coordinate frames, which restores covariance for scattering experiments \cite{schweber1962}.  (c)  In the situation of interest here, the interaction vanishes in coordinate frame $O$ at the time of the measurement but it is nonzero in $O'$.  This ensures that the operator that corresponds to the measurement is well defined in $O$, while the nonzero interaction in $O'$ invalidates the usual proof of covariance.  }
\label{Fig 3}
\end{figure}

The situation of interest here is shown in Fig.\ 3(c), where the interaction is zero at the time of the measurement in $O$ but not in $O'$.  The fact that the interaction is zero in $O$ avoids the usual concern that the operator that represents the measurement may not be well-defined in the presence of an interaction, while the fact that the interaction is nonzero in $O'$ invalidates the usual proof of covariance in Fig.\ 3(b).  

The interactions in regions A and B are spacelike separated from the measurement with $\{ \hat{\psi} (\mathbf{r}_e,t), \bar{\psi}(\mathbf{r}_S,t') \}=0$.  Nevertheless, Eq.\ \eqref{p1prime459} shows that it is the commutator of the field in the interferometer with higher-order terms in the time-evolution operator that determines the phase shifts in the interferometer, and those commutators involve the fields at two different points in the interferometer that are timelike separated, as discussed in the Introduction.  Once again, this can be intuitively understood from the fact that the action in the Feynman path integral approach is that of the entire system, including the interactions in regions A and B.

\section{CALCULATION OF THE AHARONOV-BOHM PHASE SHIFT}

The magnetic Aharonov-Bohm effect is illustrated schematically in Fig.\ 1.  The field of an incident electron is split into two paths labeled 1 and 2 by a beam splitter or diffraction.  The two paths enclose a solenoid S that generates a magnetic flux $\Phi_M$
inside the solenoid, with the electric field $\mathbf{E}=0$ and the magnetic field $\mathbf{B}=0$ outside of the solenoid.  The vector potential $\mathbf{A}$ is nonzero along paths 1 and 2, which produces a phase difference of $\Delta \phi$.  The probability of detecting the electron in the two output paths $1'$ and $2'$ of the interferometer is  proportional to $\cos^2 (\Delta \phi /2)$ in one path and $\sin^2 (\Delta \phi /2)$ in the other path.

The Aharonov-Bohm effect will be derived here by extracting the phase shift in the interferometer directly from the time-evolution operator.  The same results are obtained in Section VI and Appendix B using in-in perturbation theory.

 In quantum electrodynamics, the interaction Hamiltonian that couples the electrons and the electromagnetic field is given by
\begin{equation}
    {\hat H}'(t)=-\int d^3 \mathbf{r} \, \hat{A} ^ \mu  (\mathbf{r},t)  \hat{j}_ \mu  (\mathbf{r},t).
    \label{hprime11}
\end{equation} 
Here $\hat{A} ^ \mu$ is the second-quantized vector potential while $\hat{j}_ \mu $ is the current operator for the electrons.  The Coulomb potential $\hat{A} ^0$ can be neglected here because the solenoid is uncharged, in which case Eq.\  \eqref{hprime11} can be rewritten as
\begin{equation}
    {\hat H}'(t)=-\int d^3 \mathbf{r} \, \hat{\mathbf{A}} (\mathbf{r},t) \cdot \hat{\mathbf{j}} (\mathbf{r},t),
    \label{hprime3}
\end{equation}
where $\hat{\mathbf{A}} (\mathbf{r},t) $ and $\hat{\mathbf{j}} (\mathbf{r},t)$ are three-vectors.  The metric $g^{\mu \nu}$ will be chosen to have diagonal elements of (-1,1,1,1).

In the Schr\"odinger picture, the second-quantized vector potential  is defined in the Lorentz gauge as \cite{peskin1995}
\begin{equation}
    \hat{A}^\mu (\mathbf{r},t) = \frac{1}{(2 \pi)^3} \sum_ \lambda \int \frac{d^3 \mathbf{k}}{\sqrt{2 \omega _k}} \left[ \epsilon ^\mu (\mathbf{k}, \lambda) \hat{a} _ \lambda (\mathbf{k}) e^{i \mathbf{k} \cdot \mathbf{r}  } 
       + h.c.   \right] ,
    \label{vectorpotential}
\end{equation}
where $\hat{a} _ \lambda (\mathbf{k}) $ annihilates a photon of momentum $\mathbf{k}$ and polarization ${\epsilon ^\mu} (\mathbf{k}, \lambda) $, while $\omega_k=ck$. 

Although the motion of the electrons is nonrelativistic, it is more straightforward to use the Dirac theory where the current operator is given by $\hat{j}^ \mu=q\Bar{\psi} \gamma^ \mu \hat{\psi }$.  Here $\hat{\psi } (\mathbf{r},t)$ is the second-quantized field operator for the electrons and the $\gamma ^ \mu$ are related to the Dirac matrices.  This approach has the advantage that the current operator does not involve the vector potential and $H'$ in Eq.\ \eqref{hprime11} is exact.

Retardation effects due to the time dependence of the currents will play a major role in what follows.  As a result, the interaction Hamiltonian will be time dependent and the time-evolution operator  $\hat{U}(t_f,t_i)$ in the interaction picture is given by the Dyson formula \cite{dyson1949}
\begin{equation}
    \hat{U}(t_f,t_i)=\hat{T} \, \exp \left[ -i \int_{t_i}^{t_f} \hat{H}'(t) dt   \right] .
    \label{Dyson}
\end{equation}
Here $\hat T$ is the time-ordering operator that places operators with the earliest time to the right, while $t_i$ and $t_f$ are the initial and final times.
Combining Eqs.\ \eqref{hprime3} and \eqref{Dyson} gives
\begin{equation}
    \hat{U}= \hat{T} \, \exp \left[ i \int d^3 \mathbf{r} \,\int_{t_i}^{t_f}  dt \,    \hat{\mathbf{A}} (\mathbf{r},t) \cdot \hat{\mathbf{j}} (\mathbf{r},t)     \right] .
    \label{Dyson2}
\end{equation}

First consider the phase shift $\phi_1$ that is generated along path 1.  The interaction between the electrons is due to the exchange of virtual photons that are generated  from the vacuum state of the field, which can be calculated as usual using perturbation theory \cite{weinberg1995,peskin1995,feynman1948,feynman1950}.  The initial state of the system will be denoted by $ | \Psi_0 \rangle = | \Psi_e \rangle  | 0 \rangle$, where $| \Psi_e \rangle $ is the initial state of the electrons and $ | 0 \rangle$ is the vacuum state of the electromagnetic field.

 The interaction  will produce a final state $ |\Psi _f \rangle =\hat{U} |\Psi _0 \rangle$. There will be  a small probability amplitude to produce a state that is orthogonal to $|\Psi _0 \rangle $, which would reduce the visibility of the interference pattern, and a larger probability amplitude to remain in the initial state with an overall phase shift given by $|\Psi _f \rangle =\exp(i \phi_1) |\Psi _0 \rangle$. Thus the phase shift $\phi_1$ is  given by
 \begin{equation}
     e^{i \phi_1 }= \langle \Psi _o | \hat{U}(\Delta t) | \Psi _ 0 \rangle, 
     \label{projection}
 \end{equation}
 where $\Delta t = t_f - t_i$.
 Berry used a similar approach to derive his  geometric phase \cite{berry1984}.  The same results are derived in a more formal way using in-in perturbation theory in Section VI and Appendix B. 

 Perturbation theory is equivalent to expanding the exponential in Eq.\ \eqref{Dyson2} in a Taylor series, which can be used to derive the Feynman diagrams \cite{dyson1949}.  Expanding Eq.\ \eqref{Dyson2}  to second order and inserting it into Eq.\ \eqref{projection} gives
 \begin{eqnarray}
 e^{i \phi_1} &=& \langle \Psi_0 |\Psi_0 \rangle \nonumber \\[2pt]
     &+& i   \langle \Psi_0 | \int d^3 \mathbf{r} \, \int_{t_i}^{t_f}  dt \,    \hat{\mathbf{A}} (\mathbf{r},t) \cdot \hat{\mathbf{j}} (\mathbf{r},t) |\Psi_0 \rangle 
 \nonumber\\[2pt]
 &-& \frac{1}{2}\langle \Psi_0 | \int d^3 \mathbf{r} \int d^3 \mathbf{r}' \,\int_{t_i}^{t_f}  dt \int_{t_i}^{t_f}  dt' \nonumber\\[2pt]
     & \times &  \hat{T} \, \left[ \hat{\mathbf{A}} (\mathbf{r},t) \cdot \hat{\mathbf{j}} (\mathbf{r},t) \,   \hat{\mathbf{A}} (\mathbf{r'},t') \cdot \hat{\mathbf{j}} (\mathbf{r'},t') \right] |\Psi_0 \rangle .
 \label{expansion}
 \end{eqnarray}
 The first term in Eq.\ \eqref{expansion} is unity and the second term is zero because $\langle 0 | \hat{a} |0  \rangle =  \langle 0 | \hat{a}^\dagger |0  \rangle =0$.
 It can be seen that the Taylor series expansion of the exponential has introduced a factor of 1/2 in the final term, which is necessary to eliminate any double-counting of the interaction.  Equation \eqref{expansion} is equivalent to the Feynman diagrams shown in Fig. 2.

 Consider the case in which $\mathbf{r}$ is located in one of the paths through the interferometer while $\mathbf{r}'$ is located in the solenoid.  In that case,  $\hat{\mathbf{j}} ({\mathbf{r}},t)$ and $\hat{\mathbf{j}} ({\mathbf{r'}},t')$ are independent operators and $[\hat{\mathbf{j}} ({\mathbf{r}},t),\hat{\mathbf{j}} ({\mathbf{r'}},t')]=0$ in the limit of large separations. (This is equivalent to neglecting the Feynman propagator for an electron, which decreases rapidly with distance.)  Thus the time ordering of the current operators is irrelevant and 
 \begin{eqnarray}
     \hat{T}  \left[ \hat{\mathbf{A}} (\mathbf{r},t) \cdot \hat{\mathbf{j}} (\mathbf{r},t) \,   \hat{\mathbf{A}} (\mathbf{r'},t') \cdot \hat{\mathbf{j}} (\mathbf{r'},t') \right] \nonumber \\[2pt]
    = \hat{j}_i (\mathbf{r},t) \hat{T}  \left[ \hat{{A}^i} (\mathbf{r},t)    \hat{{A}^k} (\mathbf{r'},t')   \right] \hat{j}_k (\mathbf{r'},t').
    \label{neworder} 
 \end{eqnarray}
 
 By definition, the Feynman propagator $\Delta _F ^{\mu \nu}$ for a photon is given by \cite{Feynman1949,feynman1950,weinberg1995,peskin1995,griffiths2008,schwartz2014,dickinson2014}
 \begin{equation}
     \Delta _F ^{\mu \nu}  (\mathbf{r},t;  \mathbf{r'},t')  
     =\langle 0  \, | \hat{T}  \left[ \hat{A}^\mu (\mathbf{r},t)    \hat{A}^\nu (\mathbf{r'},t')  \right] |  0 \rangle .  
     \label{Feynman}
 \end{equation}
 (Other conventions may differ by a factor of i.)  Inserting Eqs.\ \eqref{neworder} and \eqref{Feynman} into Eq.\ \eqref{expansion} gives
 \begin{eqnarray}
     i \phi_1 & = & - \frac{1}{2}\ \int d^3 \mathbf{r} \int d^3 \mathbf{r}' \,\int_{t_i}^{t_f}  dt \int_{t_i}^{t_f}  dt' \nonumber \\[2pt]
     & \times & \langle \Psi_e | \hat{j}_i (\mathbf{r},t)  \, \Delta _F ^{i k} (\mathbf{r},t;  \mathbf{r'},t')   \hat{j}_k (\mathbf{r'},t') |  \Psi_e \rangle .
     \label{result1}
 \end{eqnarray}
 Here it has been assumed that the interaction is sufficiently weak that $\exp (i \phi_1) \simeq 1+i \phi_1$.  The process can be divided into smaller time intervals if necessary in order to satisfy that approximation. 

It will also be assumed that the interaction is sufficiently weak that it does not generate any significant correlation between the two currents. In that case,
\begin{equation}
     \langle  \hat{j}_i (\mathbf{r},t) \hat{j}_k  (\mathbf{r}',t')  \rangle \\  
    = \langle \hat{j}_i (\mathbf{r},t)  \rangle \, \langle  \hat{j}_k  
    (\mathbf{r}',t')  \rangle.
    \label{jj}
    \end{equation}
The expectation value of the current will be denoted by ${J}_i ({r},t)=\langle \Psi_e | \hat{j}_i (\mathbf{r},t) | \Psi_e  \rangle $.  Inserting this into Eq.\ \eqref{result1} gives
 \begin{eqnarray}
     i \phi_1 & = & - \frac{1}{2}\ \int d^3 \mathbf{r} \int d^3 \mathbf{r}' \,\int_{t_i}^{t_f}  dt \int_{t_i}^{t_f}  dt' \nonumber \\[2pt]
     & \times &  J_i (\mathbf{r},t)  \, \Delta _F ^{i k}(\mathbf{r},t;  \mathbf{r'},t')   J_k (\mathbf{r'},t'). 
     \label{result6}
 \end{eqnarray}

  If the Coulomb potential had been included in $H'$ in Eq.\ \eqref{hprime3}, then Eq.\ \eqref{result6} could be written in the form
 \begin{equation}
  i  \phi_1 = - \frac{1}{2}\ \int d^4 {x} \int d^4 {x}' \,  J_\mu ({x})  \, \Delta _F ^{\mu \nu} ({x}-  {x'})   J_ \nu ({x'}),
     \label{result17}  
 \end{equation}
 where $x=(\mathbf{r},t)$.  Despite the covariant form of the integrand, Eq.\ \eqref{result17} is not covariant unless $t_f \rightarrow \infty$ for the reasons discussed in Section II.  The same situation would occur using Feynman path integrals, since the Lagrangian density is invariant but the action is not unless the integrals are extended to $t_f \rightarrow \infty$.

 The Feynman propagator is a function of $x-x'$  and it obeys the relation \cite{Jauch1980,dickinson2014}
 \begin{eqnarray}
     \Delta _F ^{\mu \nu} (x-x') &=&\frac{1}{2} \left[ \Delta_R^{\mu \nu} (x-x')+\Delta_A^{\mu \nu} (x-x') \right] \nonumber \\[2pt]
    & +& \frac{1}{2} \Delta_1^{\mu \nu} (x-x'). 
    \label{advancedretarded} 
 \end{eqnarray} \\[2pt]
 The retarded propagator $\Delta_R^{\mu \nu}$ is related to the retarded Green's function $G_R^{\mu \nu}$ of classical electrodynamics by 
 \begin{equation}
     \Delta_R^{\mu \nu} (x-x')
     = - i G_R^{\mu \nu} (x-x').  
     \label{Greensfn} 
 \end{equation} 
 A similar relation holds for the advanced propagator $\Delta_A^{\mu \nu}$.  The Hadamard function $\Delta_1^{\mu \nu} $  is real and it does not contribute to the phase shift to lowest order, which allows it to be neglected here. ($\Delta_R$ and $\Delta_A$ are imaginary in the convention of Eq.\ \eqref{Feynman}.)

The integral over $dt'$ in Eq.\ \eqref{result6} can be divided into two intervals to give
\begin{eqnarray}
     && \int_{t_i}^{t_f}  dt \int_{t_i}^{t_f}  dt'   J_i (\mathbf{r},t)  \, \Delta_F^{i k} (\mathbf{r},t;  \mathbf{r'},t')   J_k (\mathbf{r'},t') \nonumber \\[2pt] 
      &=&
       \int_{t_i}^{t_f}  dt \int_{t_i}^{t}  dt' \, \,   J_i (\mathbf{r},t)  \, \Delta_F^{i k} (\mathbf{r},t;  \mathbf{r'},t')   J_k (\mathbf{r'},t') \nonumber \\[2pt] 
       &+&
       \int_{t_i}^{t_f}  dt \int_{t}^{t_f}  dt'   J_i (\mathbf{r},t)  \, \Delta_F^{i k} (\mathbf{r},t;  \mathbf{r'},t')   J_k (\mathbf{r'},t').
     \label{newintegral}
 \end{eqnarray}
 Interchanging the order of integration in the last term of Eq.\ \eqref{newintegral} and making the change of variables $ \mathbf{r},t \leftrightarrow \mathbf{r}',t'$ gives
 \begin{eqnarray}
     && \int_{t_i}^{t_f}  dt \int_{t_i}^{t_f}  dt'   J_i (\mathbf{r},t)  \, \Delta_F^{i k} (\mathbf{r},t;  \mathbf{r'},t')   J_k (\mathbf{r'},t') \nonumber \\[2pt] 
      &=&
       \int_ {t_i} ^ {t_f}  dt \int_ {t_i} ^ {t}  dt'   J_i (\mathbf{r},t)  \, \Delta_R^{i k} (\mathbf{r},t;  \mathbf{r'},t')   J_k (\mathbf{r'},t') ,
     \label{newintegral2}
 \end{eqnarray}
where  Eq.\ \eqref{advancedretarded} was used along with the fact that $\Delta _F ^{\mu \nu}(\mathbf{r},t)$ is an even function of $\mathbf{r}$ and $t$.  Inserting Eqs.\ \eqref{Greensfn} and \eqref{newintegral2} into Eq.\ \eqref{result6} gives 
 \begin{eqnarray}
     \phi_1 & = & \frac{1}{2}\ \int d^3 \mathbf{r} \int d^3 \mathbf{r}' \,\int_{t_i}^{t_f}  dt \int_{t_i}^{t}  dt' \nonumber \\[2pt]
     & \times &  J_i (\mathbf{r},t)  \, G_R^{ik} (\mathbf{r},t;  \mathbf{r'},t')   J_k (\mathbf{r'},t'). 
     \label{result7}
 \end{eqnarray}

 Taking $t_i \rightarrow -\infty $, the integrals over $d^3 \mathbf{r}'$ and $dt'$ give
 \begin{eqnarray}
     \int d^3 \mathbf{r}' \int _{-\infty} ^ t   dt'   G_R^{ik} (\mathbf{r},t;  \mathbf{r'},t')   J_k (\mathbf{r'},t')=\mathcal{A}^{i} (\mathbf{r},t), 
     \label{result8}
 \end{eqnarray}
 where $\boldsymbol{\mathcal{A}} (\mathbf{r},t)$ is the expectation value of the retarded vector potential produced by the current distributions.  Combining this with Eq.\ \eqref{result7} gives
 \begin{equation}
     \phi_1  =  \frac{1}{2}\ \int d^3  \mathbf{r}  \int _{-\infty} ^ {t_f} dt  \,
     \boldsymbol{{J}}  (\mathbf{r},t) \cdot \boldsymbol{\mathcal{A}} (\mathbf{r},t).
     \label{result10}
 \end{equation}

The expectation value of the total current can be written in the form
\begin{equation}
\boldsymbol{{J}}=\boldsymbol{{J}}_e+{\boldsymbol{J}}_S
    \label{jsum},
\end{equation}
where ${\boldsymbol{J}}_e$ is the expectation value of the current produced by the electron in the interferometer while $\boldsymbol{J}_S$ is the expectation value of the current in the solenoid. In the same way, the expectation value of the total vector potential can be written as
\begin{equation}
\boldsymbol{\mathcal{A}}= \boldsymbol{\mathcal{A}}_e+ \boldsymbol{\mathcal{A}}_S,
    \label{Asum}
\end{equation}
where $ \boldsymbol{\mathcal{A}}_e $ and $ \boldsymbol{\mathcal{A}}_S $ are the expectation values of the retarded vector potentials produced by the electrons in the interferometer and in the solenoid, respectively. Inserting Eqs.\ \eqref{jsum} and \eqref{Asum} into Eq.\ \eqref{result10} gives
 \begin{eqnarray}
   \phi_1 =  \frac{1}{2}\ \int d^3 \mathbf{r} \, dt \, \left[ 
      \boldsymbol{J}_e   \cdot \boldsymbol{\mathcal{A}}_{S} +   \boldsymbol{J_S}   \cdot \boldsymbol{\mathcal{A}}_{e} \right] 
     \label{result12}
 \end{eqnarray}
as stated in the introduction,  where both vector potentials are retarded.  Self-interaction terms such as $ \boldsymbol{J}_e   \cdot \boldsymbol{\mathcal{A}}_{e}$ have been omitted since they are the same along both paths and do not contribute to the phase difference.  
 
 Equation \eqref{result12} is one of the main results of this paper.  It shows that the phase shift in the interferometer depends on the interaction at the solenoid as well as the usual interaction at the interferometer, as described in more detail in the following section.

 The two symmetric interaction terms in Eq.\ \eqref{result12} are similar in some respects to the usual expression for the Coulomb energy $U_C$ of a system of $N$ particles in classical electrostatics, where
 \begin{equation}
     U_C =\frac{1}{2} \sum _{i\neq j} ^N q_i \Phi _j .
     \label{electrostatics}
 \end{equation}
 The factor of 1/2 is necessary in order to avoid any double-counting of the interaction. The vector potentials in Eq.\ \eqref{result12} are retarded while the Coulomb potential in Eq.\ \eqref{electrostatics} is not.  Other than that, the two expressions are analogous and the symmetric form of Eq.\ \eqref{result12} should not be too surprising.

 The phase shift was calculated here by extracting it from the time-evolution operator using the projection onto the initial state in Eq.\ \eqref{projection}, which is the most straightforward approach.  The expectation value $\langle \Phi_0 | \hat{U} ^\dagger \hat{M} \hat{U}  | \Phi_0 \rangle $ of an operator $\hat{M}$ is often calculated using in-in (Schwinger-Keldysh) perturbation theory  \cite{schwinger1961,Keldysh1965,baym1961,baym1962,Kadanoff1962,rammer2007}), which is more formal than the approach used in the text.  It is shown in Appendix B that the use of in-in perturbation theory gives the same results as are obtained in the text.

\section{QUASISTATIC FIELDS AND THE OVERALL PHASE SHIFT}

The Aharonov-Bohm effect for the case of quasistatic fields where retardation effects are negligible will be considered in this section.  The role of the overall phase of the system will also be discussed in more detail.

In the Schr\"odinger picture, the initial state $|\psi_i \rangle$ of the system before the first beam splitter has the form
\begin{equation}
    |\psi_i \rangle = |S_i \rangle |0\rangle |P_i \rangle.
    \label{initialstate}
\end{equation}
Here $|S_i \rangle$ is the initial state of the source while $|P_i \rangle$ is the state of the electron in the incident path $P_i$ to the interferometer.  The electromagnetic field is assumed to initially be in the  vacuum state $|0\rangle$ for $t_i \rightarrow - \infty.$  (Virtual photons associated with the Coulomb field can be neglected here.)

The first beam splitter is assumed to produce an equal probability amplitude for the electron to be in paths 1 or 2, so that the state of the system just before the final beam splitter can be written as
\begin{equation}
    |\psi \rangle = \frac{1}{\sqrt{2} } \left[ \, e^{i \phi _1} |S_1 \rangle |F_1\rangle |P_1 \rangle  +  e^{i \phi _2} |S_2 \rangle |F_2\rangle |P_2 \rangle  \right].
    \label{middlestate}
\end{equation}
Here $|S_1 \rangle$,  $|F_1 \rangle$, and $|P_1\rangle$ are the final states of the source, electromagnetic field, and electron after the interaction for the case in which the electron took path 1 through the interferometer.  The states $|S_2 \rangle$,  $|F_2 \rangle$, and $|P_2\rangle$   correspond to the case in which the electron traveled along path 2.  Any phase shifts due to the interaction have been included in $\phi_1 $ and $\phi_2.$   The states $|F_1\rangle $ and $|F_2\rangle $ include the possibility that an infrared photon may have been radiated by the  electron, while any virtual photons will vanish at the end of the process when the currents go to zero.  

The product of states $|S_2 \rangle |F_2\rangle$ in Eq.\ \eqref{middlestate} can be written as its projection onto $|S_1 \rangle |F_1\rangle$  with an additional orthogonal term $| \psi_\perp \rangle$, which gives
\begin{equation}
    |S_2 \rangle |F_2\rangle   = \chi  |S_1 \rangle |F_1\rangle  +\xi | \psi_\perp \rangle.
    \label{proj3}
\end{equation}
The  coefficients $\chi$ and $\xi$ can be taken to be real by absorbing any phase difference into $\phi_1$ and $\phi_2$.  Inserting Eq.\ \eqref{proj3} into Eq.\ \eqref{middlestate} gives
\begin{eqnarray}
    |\psi \rangle &=& \frac{1}{\sqrt{2} }|S_1 \rangle |F_1\rangle    \left[ \, e^{i \phi _1}|P_1 \rangle  + \chi e^{i \phi _2} |P_2 \rangle \right] \nonumber \\[2pt]
    &+& \frac{1}{\sqrt{2} } \xi e^{i \phi _2} | \psi_\perp \rangle |P_2 \rangle  .
    \label{withproj}
\end{eqnarray}

The final beam splitter couples $|P_1 \rangle$ and $|P_2 \rangle$ into output path $1'$ to give a final state $|\psi_{1'} \rangle$ of the form
\begin{eqnarray}
    |\psi_{1'} \rangle &=& \frac{1}{2 }|S_1 \rangle |F_1\rangle |P_{1'} \rangle    \left[ \, e^{i \phi _1} + \chi e^{i \phi _2}  \right] \nonumber \\[2pt]
    &+& \frac{1}{2 } \xi  e^{i \phi _2} | \psi_\perp \rangle |P_{1'} \rangle,  
    \label{finaltext}
\end{eqnarray}
where $|P_{1'} \rangle$ denotes a state with the electron in output path $1 ' $.  A similar expression exists for the case in which the electron is in output path $2 ' $.

 In the limit of a weak interaction, the change in the state of the source will be small and $|S_1 \rangle \approx |S_2 \rangle \approx |S_i \rangle$, where any phase shifts due to the interaction are once again included in $\phi_1$ or $\phi_2$.   In a similar way, the effects of infrared photon emission can be shown to be negligible under the usual experimental conditions \cite{Stern1990,Ford1993,Breuer2001,delisle2024}, which is equivalent to $|F_1 \rangle \approx |F_2 \rangle \approx |0 \rangle$.  In that case $\chi \approx 1$ and $\xi \approx 0,$ and the state of the system for an electron in path $1'$ after the final beam splitter can be written to a good approximation as 
\begin{equation}
    |\psi_{1'} \rangle = \frac{1}{2 } |S_i \rangle |0\rangle|P_1 ' \rangle    \left[ \, e^{i \phi _1} +  e^{i \phi _2}   \right].
    \label{finalstate}
\end{equation}
Taking the norm of this expression gives the probability of detecting the electron in path $1'$, which is proportional to $\cos^2[(\phi_1-\phi_2)/2]$.  A nonzero value of $\chi$ would reduce the visibility of the interference pattern. The effects of entanglement between the electron in the interferometer and the final state of the source are discussed in more detail in  Appendix A.

It can be seen from Eq.\ \eqref{middlestate}  that any phase shifts originating in $|S_1 \rangle,$  $|F_1\rangle,$ or $ |P_1 \rangle$ can all contribute to the overall phase $\phi_1$, with a similar set of contributions to $\phi_2$.  This explains why the output of the interferometer can depend on the $\boldsymbol{J}_S \cdot \boldsymbol{\mathcal{A}}_e $ interaction at the source of the magnetic field as well as the usual  $\boldsymbol{J}_e \cdot \boldsymbol{\mathcal{A}}_S $  interaction at the interferometer.  Both of these interactions can contribute to the total energy of the system and thus its phase. 

The dependence on the overall phase of the system can  be further understood by considering the use of Feynman path integrals.  In that case, the probability amplitude for each path in the  configuration space of the system is given by $e^{i \mathcal{S}}$, where $\mathcal{S}$ is the action.  It follows from this that the phase associated with each such path is the total phase shift of the entire system, in agreement with the results obtained above.  The dependence of the interferometer output on the overall phase of the system is also derived in Appendix B using in-in perturbation theory.

It may seem counterintuitive that the output of an electron interferometer can depend on a phase shift that occurs in a distant source.  This situation is analogous to the nonlocal interference between two entangled photons that pass through two distant interferometers, where the output of one interferometer depends on the phase settings in both interferometers \cite{franson1989,Gisin1998}. Nonlocal interferometers of that kind violate Bell's inequality but not causality, and they are widely used in quantum communication protocols \cite{ribordy2000,brougham2013,bulla2023,fallon2025}.  The nonlocal nature of the overall phase of the system is the essential feature in both situations.

One might ask whether or not there is an additional phase shift associated with the electromagnetic field that could cancel out the effects of interest here.  That is not the case, since the time-evolution operator in Eq.\ \eqref{Dyson} gives the total phase shift of the entire system including that of the electrons and the electromagnetic field.

 For quasistatic fields where retardation effects are negligible, it can be shown that
\begin{equation}
    \int d^3 \mathbf{r} \,dt \, \boldsymbol{J}_S \cdot \boldsymbol{\mathcal{A}}_e =  \int d^3 \boldsymbol{r} \, dt \, \boldsymbol{J}_e \cdot \boldsymbol{\mathcal{A}}_S .
    \label{phi9}
\end{equation}
In that case, Eq.\ \eqref{result12} can be rewritten as
\begin{eqnarray}
   \phi_1 =   \int d^3 \mathbf{r} \, dt \, 
      \boldsymbol{J}_e   \cdot \boldsymbol{\mathcal{A}}_{S}  .
     \label{result18}
 \end{eqnarray}
 If the electron in the interferometer is described by a localized wave packet, the spatial integral of the current density gives $q\mathbf{v}$ in the nonrelativistic limit, where $\mathbf{v}$ is the velocity.  Taking the line integral around the two paths gives  
\begin{equation}
    \Delta \phi = \frac{q}{ \hbar} \oint {\boldsymbol{\mathcal{A}}_S \cdot\ d  \mathbf{l}},
    \label{lineintegral2}
\end{equation}
which is the usual expression for the AB effect. This is equivalent to  quantizing only $\hat{\mathbf{j}}_e$ in a conventional approach based on a classical vector potential.  
 
Equation \eqref{phi9} could be used equally well to rewrite Eq.\  \eqref{result12} as 
\begin{eqnarray}
   \phi_1 =   \int d^3 \mathbf{r} \, dt \, 
      \boldsymbol{J}_S   \cdot \boldsymbol{\mathcal{A}}_{e}  
     \label{result19}
 \end{eqnarray}
 for quasistatic fields.  This shows that the phase shift could be obtained by quantizing  the electrons in the solenoid instead in agreement with Ref.\ \cite{vaidman2012role}.

 These results show that the predictions of quantum field theory are in agreement with the experimentally-observed value of the magnetic Aharonov-Bohm effect in the quasistatic limit where retardation effects are negligible.

\section{RETARDED POTENTIALS AND CAUSALITY}

 The  results obtained above for the magnetic Aharonov-Bohm effect are more problematic when retardation effects are significant.  In that case, conventional field theory predicts a fractional Aharonov-Bohm effect, a lack of covariance, and violations of causality, as discussed in more detail below.  
 
 First consider the situation illustrated in Fig.\ 4, where a constant current in source S generates a static magnetic field.  The distance $D$ between the source and an electron interferometer is assumed to be sufficiently large that $c \Delta t<<D$, where $\Delta t$ is the transit time of an electron through the interferometer.  
 
\begin{figure}[tbp]
\centering
\includegraphics[width=0.35\textwidth]{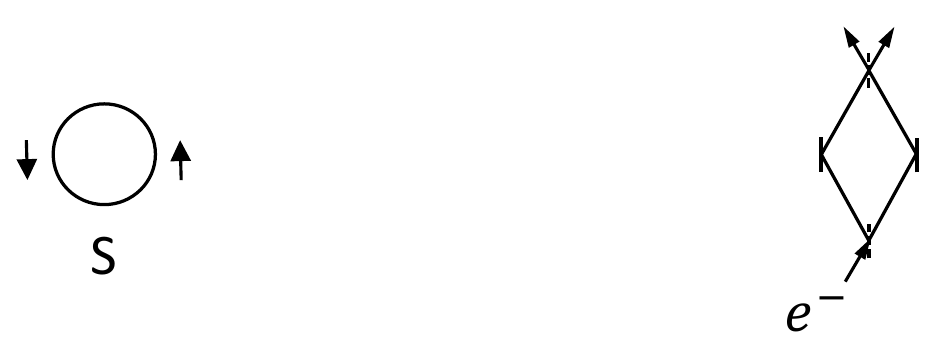}
\qquad
\caption{Fractional Aharonov-Bohm effect.  A source S of a static magnetic field consists of a constant current flowing in a circular loop.  An electron passes through a distant interferometer, which encloses a magnetic   flux from the source.  The distance D between the source and the interferometer is assumed to be sufficiently large that the electron passes through the interferometer before its retarded vector potential can reach the source S.  In that case, Eq. \eqref{phi3} predicts that a fractional Aharonov-Bohm phase shift will be observed. }
\label{Fig 4}
\end{figure}

 It will also be assumed that the current $\boldsymbol{J}_e$ in the interferometer is zero outside of the time interval $\Delta t$.  As a result, the retarded vector potential $\boldsymbol{\mathcal{A}}_{e} (\mathbf{r},t)$ at the source S is zero throughout the entire process and the last term in Eq.\ \eqref{result12} is zero.  Reinserting factors of $\hbar$, the remaining term gives an observable phase difference between the two paths of
\begin{equation}
    \Delta \phi = \frac{1}{2} \frac{q}{\hbar } \oint {\boldsymbol{\mathcal{A}}_S \cdot\ d  \mathbf{l}}.
    \label{phi3}
\end{equation}\\
Equation \eqref{phi3} predicts that a fractional AB effect can occur as a result of retardation effects.  A fractional AB effect has previously been observed in solid-state systems for reasons that are unrelated to retardation. \cite{kusmartsev1994,emperador2003,nakamura2019}.

A number of earlier papers \cite{franson1995,Pearle2017b,Vedral2020,saldanha2021,Wakamatsu2024} also quantized the electromagnetic field using different techniques.  But those authors did not consider the effects of retarded potentials and none of them predicted a fractional AB phase shift with the exception of Ref.\ \cite{franson1995}.  The approach in that paper would have double-counted the interaction and given an incorrect result were it not for cancellation from a sign error.

It may be possible to test these predictions using a superconducting quantum interference device (SQUID), which is a magnetometer with a loop  geometry similar to Fig. 1 and a response time  as short as $1 \mu s$ \cite{trepanier2013realization}.  Light can travel a distance of 300 m during that time, and the separation between the source and the SQUID would have to be larger than that to eliminate the retarded vector potential at the source.  An effect of that kind would have practical implications for high-precision applications of SQUIDs if it is physically correct, but the example considered below suggests that it is an incorrect prediction due to the use of the Schr\"odinger equation in quantum field theory.

This can be seen in the Aharonov-Bohm experiment illustrated in the spacetime diagram of Fig.\ 5, where an experimenter can control a time-dependent current $J_c (t)$ in the source of a magnetic field after the time $t_{1/2}$.  The distance between the source and the interferometer is assumed to be sufficiently large that the retarded vector potential produced by $J_c (t)$ cannot reach the interferometer before the electron reaches the final beam splitter and the output path is measured.  In that case, $J_c (t)$ and the measurement process are spacelike separated.  But according to Eq.\ \eqref{result12}, an experimenter could control the output of the interferometer by changing the current $J_c (t)$, which would allow signals to be transmitted outside of the forward light cone in violation of causality. 

\begin{figure}[tbp]
\centering
\includegraphics[width=0.34\textwidth]{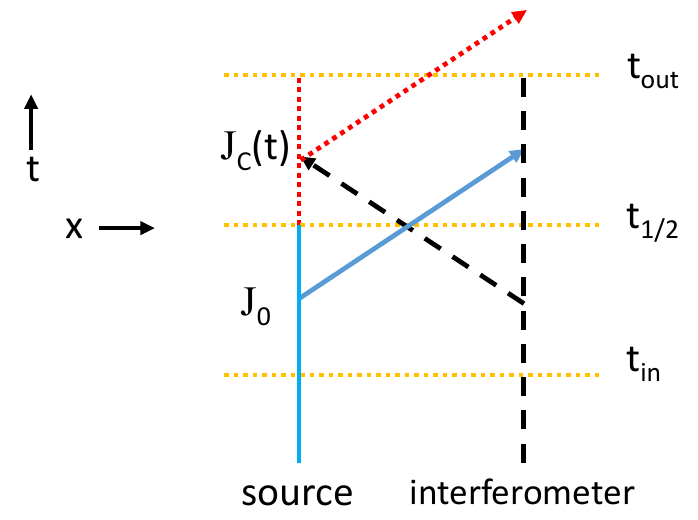}
\qquad
\caption{Spacetime diagram of an Aharonov-Bohm experiment with a time-dependent current $J_C (t)$ in the source of a magnetic field that can be controlled by an experimenter. (The geometry is shown in Fig.\ 4).  A single electron enters an interferometer at time $t_{in}$ and emerges from the second beam splitter at time $t_{out}$, when the output path taken by the electron is measured.  The time $t_{1/2} = (t_{in} + t_{out} )/2$. The current in the source is assumed to have a constant value $J_0$ until time $t_{1/2}$, after which it  has a time-dependent value $J_C (t)$. The currents associated with the electron in the interferometer and the source are both assumed to be zero after time $t_{out}$ in coordinate frame O.  The distance between the source and the interferometer is chosen in such a way that the transit time between them at the speed of light is  $t_{1/2}-t_{in}$.  Thus the vector potential generated by $J_C (t)$ cannot reach the interferometer by the time of the measurement, and the control current is spacelike separated from the measurement process.  Nevertheless, Eq.\ \eqref{result12} predicts that $J_C (t)$ can have an effect on the output of the interferometer in violation of causality.      }
\label{Fig 5}
\end{figure}

All of the currents are assumed to be zero after time $t_{out}$ in the coordinate frame $O$ shown in Fig. 5.  Since the two currents are spacelike separated at time $t_{out}$, there are other coordinate frames $O'$ where $J_c (t)$ is applied after the measurement has already been made.  Thus Eq.\ \eqref{result12} would give different results in two different reference frames which shows that the theory is not covariant, as expected from the argument shown in Fig. 3.

There may be some concern that these difficulties could be due to a limitation in the use of perturbation theory, as suggested by Haag's theorem \cite{Haag1955} for example.  That is not the case since similar results were obtained in Ref.\ \cite{franson1995} using nonperturbative techniques from quantum optics.

Equation \eqref{projection} is equivalent to taking the projection $|0\rangle \langle 0 |$ onto the vacuum state of the electromagnetic field, and it is important to consider any possible contribution from final states that are orthogonal to $ |0 \rangle$.  It is shown in Appendix B that there is no contribution to the phase shift in the interferometer from final states containing one or more photons to order $q^2$.

These results show that the use of quantum field theory gives results that are not covariant and violate causality.  The origin of these difficulties will be discussed in more detail in the next section.

\section{ROLE OF THE COMMUTATION RELATIONS IN CAUSALITY} 

The commutators or anticommutators of the field operators vanish outside of the forward light cone in the Heisenberg picture, which is generally assumed to prove that causality must hold in disagreement with the results obtained above.  This discrepancy is analyzed  in the interaction picture using in-in perturbation theory \cite{schwinger1961,Keldysh1965,baym1961,baym1962,Kadanoff1962,rammer2007}, where it is shown that the output of the interferometer depends on the commutator of the field with higher-order terms in the time-evolution operator, which can be nonzero even for spacelike separated events. 

The output of the interferometer is determined by the phase shifts in paths 1 and 2.  Consider the expectation value of the field in path 1 through the interferometer, which is given by
 \begin{eqnarray} 
\langle \Phi(t) | \hat{\psi} (\mathbf{r}_1,t) | \Phi(t) \rangle &=& \langle \Phi_0 | \hat{U}^\dagger \hat{\psi} (\mathbf{r}_1,t) \hat{U} | \Phi_0 \rangle 
 .  
    \label{p1prime38t}
\end{eqnarray}
Here $|\Phi_0 \rangle$ is the initial state of the system in the interaction picture and $\hat{\psi} (\mathbf{r}_1,t)$ is the free-space (unperturbed) field operator.

The factor of $\hat{\psi} (\mathbf{r}_1,t)   \hat{U} $ in Eq. \eqref{p1prime38t} can be rewritten as
\begin{equation}
    \hat{\psi} (\mathbf{r}_1,t)   \hat{U}= \hat{U} \hat{\psi} (\mathbf{r}_1,t)  +[\hat{\psi} (\mathbf{r}_1,t), \hat{U}]
    \label{commutatorft}.
\end{equation}
Inserting this into Eq.\ \eqref{p1prime38t} gives
 \begin{eqnarray} 
 \hspace{-0.7 cm}
\langle \hat{\psi} (\mathbf{r}_1,t) \rangle
  &=&  \langle \Phi_0 | \hat{U}^\dagger (  \hat{U} \hat{\psi} (\mathbf{r}_1,t)  +[\hat{\psi} (\mathbf{r}_1,t), \hat{U}]) | \Phi_{0} \rangle .
    \label{p1prime499t}
\end{eqnarray}
Since $\hat{U}^\dagger   \hat{U} = \hat{I}$, the first term in Eq. \eqref{p1prime499t} is independent of the interaction and gives the expectation value of the unperturbed field $\hat{\psi} (\mathbf{r}_1,t) .$ The second term gives the contribution $\langle \hat{\psi} (\mathbf{r}_1,t) \rangle _I$ from the interaction in the form
 \begin{eqnarray} 
\langle \hat{\psi} (\mathbf{r}_1,t) \rangle _I
  &=& \langle \Phi _0 | \hat{U}^\dagger [\hat{\psi} (\mathbf{r}_1,t), \hat{U}] | \Phi_{0} \rangle .
    \label{p1prime45t}
\end{eqnarray}

To second order in the interaction, Eq.\ \eqref{p1prime45t} can be written as
  \begin{eqnarray}
 \langle \hat{\psi} (\mathbf{r}_1,t) \rangle _I  &=&
     \langle  \Phi_0| \hat{U}^{\dagger(0)} \, [ \hat{\psi} (\mathbf{r}_1,t), \hat{{U}}^{(2)}  ] | \Phi_{0} \rangle \nonumber \\[2pt]      
     &+&\langle  \Phi_0| \hat{U}^{\dagger(1)} \,[ \hat{\psi} (\mathbf{r}_1,t) , \hat{{U} ^{(1)}}  ] | \Phi_{0}
     \rangle \nonumber \\[2pt] 
     &+&\langle  \Phi_0 | \hat{U}^{\dagger(2)} \,[ \hat{\psi} (\mathbf{r}_1,t), \hat{{U} ^{(0)}}  ] | \Phi_{0} \rangle   .   \label{threetermst}
 \end{eqnarray}
 Here $\hat{U} ^{(0)}$, $\hat{U} ^{(1)}$, and $\hat{U} ^{(2)}$ denote the zero, first, and second-order terms in an expansion analogous to Eq.\ \eqref{expansion}, with a similar notation for $\hat {U} ^\dagger$.

  The last term in Eq.\ \eqref{threetermst} is zero since $\hat{\psi} (\mathbf{r}_1,t)$ commutes with $\hat{{U} ^{(0)}}=\hat{I}$.  It is shown in Appendix B that the second term is proportional to the Hadamard function $\Delta_1^{\mu \nu}(x-x')\equiv \langle 0 |\{ \hat{A}^\mu(x),\hat{A}^\nu(x')  \} | 0 \rangle$, which is a real function that does not contribute to the phase shift to second order.  This is equivalent to the $\Delta_1$ term in Eq.\ \eqref{advancedretarded} that was neglected for the same reason.
  
   The remaining term in Eq.\ \eqref{threetermst} gives 
   \begin{eqnarray}
 \langle \hat{\psi} (\mathbf{r}_1,t) \rangle _I  &=&
     \langle  \Phi_0|  \, [ \hat{\psi} (\mathbf{r}_1,t), \hat{{U}}^{(2)}  ] \, | \Phi_{0} \rangle   ,   \label{commutator1t}
 \end{eqnarray}
 as stated in the introduction.  Using the same techniques as in  the text gives
 \begin{eqnarray}
     \hat{{U}}^{(2)} & = & \frac{i}{2}\  \int d^3 \mathbf{r}' \,\int_{t_i}^{t_f} dt' \int d^3 \mathbf{r}''  \int_{t_i}^{t'} dt''   \nonumber \\[2pt]
     & \times &  \hat{j}_l (\mathbf{r'},t')  \, G_R^{lm} (\mathbf{r'},t';  \mathbf{r''},t'')   \hat{j}_m (\mathbf{r''},t''), 
     \label{result78t}
 \end{eqnarray}
 where $G_R^{lm} (\mathbf{r'},t';  \mathbf{r''},t'')$ is the retarded propagator for the electromagnetic field and the expectation value in the initial (vacuum) state of the electromagnetic field has already been taken. 
 
 The identity $[\hat{A},\hat{B} \hat{C}]=[\hat{A},\hat{B}] \hat{C}+\hat{B}[\hat{A}, \hat{C}]$ gives the commutator of $\hat{\psi} (\mathbf{r}_1,t)$ with both $\hat{j}_l (\mathbf{r'},t')$ and $\hat{j}_m (\mathbf{r''},t'')$ in Eq.\ \eqref{result78t}, which can be shown to correspond to the first and second terms in Eq.\ \eqref{result12p}, respectively. The commutators can be evaluated using the definition of the retarded propagator $S_R$ for the Dirac field, which is given by \cite{peskin1995}
 \begin{eqnarray}
     S_R(\mathbf{r}_1,t;\mathbf{r}',t')=  \Theta(t -t') \{ \hat{\psi} (\mathbf{r}_1,t),  \bar{\psi} (\mathbf{r}',t')\} 
         \label{Diracpropagatort},
 \end{eqnarray} 
 where $\Theta(t -t') $ is the unit step function. The  integral involving $S_R$ can then be  evaluated using \cite{itzykson1980} 
 \begin{equation}
     \hat{\psi} (\mathbf{r}_1,t)=\int d^3 \mathbf{r}' S_R(\mathbf{r}_1,t;\mathbf{r}',t') \gamma^0 \hat{\psi} (\mathbf{r}',t').
     \label{freefieldpropt}
 \end{equation}
 Equations \eqref{commutator1t} through \eqref{freefieldpropt} can be used to derive Eq.\ \eqref{result12p} in agreement with the results obtained in Section III, as shown in more detail in Appendix B.

Eq.\ \eqref{commutator1t} shows that the phase shift in the interferometer is determined by the commutator of $\hat{\psi} (\mathbf{r}_1,t)$ with  higher-order terms in $\hat{U}$.  The acausal process shown in Fig. 2(b) corresponds to the commutator of  $\hat{\psi} (\mathbf{r}_1,t)$  with $\hat{j}_m (\mathbf{r''},t'')$ in Eq.\ \eqref{result78t} when $\mathbf{r''}$ is in path 1 of the interferometer, in which case $\hat{j}_m (\mathbf{r''},t'')$ is timelike separated from $\hat{\psi} (\mathbf{r}_1,t)$.  This timelike commutator can be nonzero even when the current in the source and the field in the interferometer are spacelike separated with  $\{\hat{\psi}(\mathbf{r}_1,t''), \bar{\psi} (\mathbf{r}_S,t') \}=0$.  It can be shown that the same results are obtained in the Heisenberg picture, which is equivalent to the interaction picture used here.

\section{Discussion AND Summary} 

Quantum field theory is often formulated using the   Hamiltonian and the Schr\"odinger equation to describe the time evolution of the system, which is equivalent to using Feynman path integrals.  Weinberg argued that this is necessary in order for the theory to be consistent with quantum mechanics \cite{weinberg1995}.  Although the Schr\"odinger equation is not covariant, it is well known \cite{schweber1962} that scattering theory will be covariant nevertheless provided that the interaction vanishes in the initial and final states in the limit of $t \rightarrow \pm \infty.$

This paper considers a different situation in which the interaction is zero in a coordinate frame $O$ when a measurement is made but it is nonzero in regions of a different coordinate frame $O'$ due to the relative nature of simultaneity under a Lorentz transformation. The fact that the interaction is zero in $O$ avoids the usual concern that the operator representing the measurement may not be well defined in the presence of an interaction, while the fact that the interaction is nonzero in $O'$ invalidates the usual proof of covariance. 

As an example, the magnetic Aharonov-Bohm effect is derived using second-quantized field theory to describe the electromagnetic field and all the electrons, including those in the source that generates the magnetic field.  The results are in agreement with the usual expression for the Aharonov-Bohm effect in the quasistatic limit where retardation effects are negligible.  In that case, half of the phase shift is due to the $\boldsymbol{J_e}   \cdot \boldsymbol{\mathcal{A}}_{S}$ interaction of the electron in the interferometer with the vector potential produced by the source, with an equal contribution from the $\boldsymbol{J_S}   \cdot \boldsymbol{\mathcal{A}}_{e}$ interaction of the electrons in the source with the vector potential produced by the interferometer.  

Both of these interactions can contribute to the total energy of the system and thus to the total phase of the quantum state for a given path of an electron through an interferometer.  This dependence on the overall phase of the system has been demonstrated previously in nonlocal interferometer experiments that violate Bell's inequality \cite{franson1989,Gisin1998,fallon2025,ribordy2000,brougham2013,bulla2023}. 

The results of the calculations are more problematic when the retardation of the vector potential is significant.  In that case, the $\boldsymbol{J_S}   \cdot \boldsymbol{\mathcal{A}}_{e}$  interaction in the source can change the output of the interferometer even when they are spacelike separated.  As a result, conventional field theory predicts a fractional Aharonov-Bohm effect, a lack of covariance, and violations of causality.  The same results are obtained by extracting the phase shift directly from the time-evolution operator or by using in-in perturbation theory.  

The violation of causality can be understood from the fact that the field in the interferometer is determined by its commutator with the time-evolution operator, not its anticommutator with the field in the source, as shown in Eq.\ \eqref{p1prime459}.  The commutator with $\hat{{U}}^{(2)}$ can be nonzero even when the interferometer and the current in the source are spacelike separated with $\{ \hat{\psi} (\mathbf{r}_i,t), \bar{\psi}(\mathbf{r}_S,t') \}=0$.

These results suggest that quantum field theory
based on the Schr\"odinger equation or Feynman path integrals is incomplete \cite{einstein1935} in the sense that it cannot give a correct description of all observable phenomena.  Finding a theory that can will be a challenge for future research.

\begin{acknowledgments}

 The author would like to acknowledge valuable discussions with Lev Vaidman and Jonathan Zalsman regarding the role of retarded potentials.
\end{acknowledgments}

\makeatletter
\renewcommand\appendixname{APPENDIX}
\makeatother

\appendix

\setcounter{figure}{0}
\renewcommand{\thefigure}{\Alph{section}\arabic{figure}}

\section{EFFECTS OF ENTANGLEMENT}
\renewcommand{\theequation}{A\arabic{equation}}
\setcounter{equation}{0}

Entanglement between the path taken by an electron in an interferometer and the final state of the electromagnetic field or the source S can reduce the visibility of the interference, which corresponds to a value of $\chi <1$ in Eq.\ \eqref{proj3}. In principle, this could  eliminate the projection onto the initial state in Eq.\ \eqref{projection}, giving zero visibility. That does not happen under the usual experimental conditions, however, for the reasons discussed below.  

For example, suppose that the electron in the interferometer could be described by a plane wave with a definite value of its momentum, which is conserved in a virtual process.  Thus the emission or absorption of a virtual photon would leave the electron in an orthogonal final state with a different momentum.  In that case, the projection onto the initial state would be zero with $\chi =0$, which would eliminate any visibility of the interference pattern. 

This situation can be avoided if the electron is described by a wave packet corresponding to a superposition of momentum eigenstates, such as a gaussian distribution with a width $\sigma_p$.  If $\sigma_p$ is much larger than the momentum of a photon, it can be shown that the emission or absorption of the photon will leave the electron very nearly in its initial state aside from a phase factor, which corresponds to $\chi \approx 1$ and a high visibility.  The experiment in Fig. 4 implicitly assumes that the electron is localized inside the interferometer, which ensures that these conditions will be met since the typical momentum of the relevant virtual photons is on the order of $\hbar/D.$

A similar argument applies to the change in the orbital angular momentum of the electron since $\mathbf{L}=\mathbf{r} \times \mathbf{p}$. A broad distribution in linear momentum ensures a broad distribution in orbital angular momentum, so that the emission or absorption of a photon will leave the electron in its initial state aside from a phase factor.  There is no change in the spin of the electron due to the interaction Hamiltonian of Eq.\ \eqref{hprime11} in the limit of low photon momenta.

The source S of the magnetic field is also assumed to have a well-defined current distribution, which again requires that the electrons in the source be in a localized superposition state with a broad distribution of linear and angular momentum.  As before, the emission or absorption of a photon will leave the source approximately in its initial state aside from a phase shift, with $\chi =1.$  

Conservation of momentum in the intermediate state is apparent when using old-fashioned time-dependent perturbation theory \cite{baym1973}, but it is not as apparent when using the Feynman propagator in Eq.\ \eqref{result6}.  There it is tacitly assumed that the expectation value of the current distribution $\mathbf{J} (\mathbf{r},t)$ is a well-defined function of the position, which requires that the electrons be localized.  Once again, this means that the emission or absorption of a photon would leave the source in very nearly its initial state aside from a phase shift.  Thus the assumption of a well-defined current distribution is sufficient to give a large projection onto the initial state.

It should be noted that the assumption that the electrons are in a localized superposition state is not equivalent to treating them classically, which would neglect the phase shifts associated with them.  The approach used here treats all of the particles or fields quantum-mechanically.

Equation \eqref{middlestate} assumes a product state for each path through the interferometer, while entanglement along a given path could produce a state with a more general form.  But entanglement of that kind is negligible in the limit of a weak interaction and it can be neglected here.  

The ability of an experimenter to control the current in the source would correspond to a time-dependent unperturbed Hamiltonian $H_0 (t)$.  It can be shown that the Dyson formula in Eq.\ \eqref{Dyson} is still valid in that case.  Moreover, the time-dependent part of $H_0 (t)$ would commute with the vector potential operator, so that the Feynman propagator for the photon would have the same properties as in free space.

The measurement to determine which output path the electron is in could be delayed from $t_{out}$ until a later time $t_f$ when the interaction is zero in both coordinate frames, which would ensure that the results of the latter measurement would be covariant as in scattering theory. It may seem reasonable to assume that the electron would have been found in the same path whether the measurement was made at $t_{out}$ or $t_f$, which suggests that a measurement at $t_{out}$ should be covariant as well. But the validity of that assumption is not apparent in a quantum field theory based on the Schr\"odinger equation and the results obtained here provide a counterexample.  Thus it is necessary to calculate the predictions of the theory for a measurement at the actual time of interest. 

\section{DERIVATION USING IN-IN PERTURBATION THEORY}
\renewcommand{\theequation}{B\arabic{equation}}
\setcounter{equation}{0}

 The expectation value $\langle \Phi_0 | \hat{U} ^\dagger \hat{M} \hat{U}  | \Phi_0 \rangle $ of an operator $\hat{M}$ is often calculated using in-in (Schwinger-Keldysh) perturbation theory  \cite{schwinger1961,Keldysh1965,baym1961,baym1962,Kadanoff1962,rammer2007} in the interaction picture. It is shown here that in-in perturbation theory gives the same results as are obtained in the text where the phase shift was extracted directly from the time-evolution operator in Eq.\ \eqref{projection}. The violation of causality observed in the text is also analyzed in more detail based on the relevant commutation relations.

 In the interaction picture, the probability $P_{1'}$ of finding an electron in output path $1'$ of the interferometer is given by 
 \begin{eqnarray}
 P_{1'}  =
\langle \Phi (t) | \int_{R_{i}} d \mathbf{r}_e ^3  \, \bar{\psi} (\mathbf{r}_e,t) 
  \hat{\psi} (\mathbf{r}_e,t) | \Phi (t) \rangle 
    \label{p1prime492} ,
\end{eqnarray} 
where $\bar{\psi} (\mathbf{r}_e,t)=\hat{\psi} (\mathbf{r}_e,t) \gamma^0$ is the four-component Dirac adjoint operator.  Both $\bar{\psi} (\mathbf{r}_e,t)$ and $\hat{\psi} (\mathbf{r}_e,t)$ are the free-space (unperturbed) field operators and  $|\Phi (t) \rangle$ is the state of the system in the interaction picture.  

The phase of the field $\hat{\psi} (\mathbf{r}_i,t)$ in each path $i$ determines the output of the interferometer, and it is instructive to first calculate $ \langle\hat{\psi} (\mathbf{r}_1,t) \rangle $ in path 1 and determine its phase shift. The effects of the beam splitter and the contribution from the other path will be included later.

The expectation value of $\hat{\psi} (\mathbf{r}_1,t)$ at time $t$ can be evaluated using the initial state $|\Phi_0 \rangle$ and the time-evolution operator $\hat{U}(t,t')$ in the interaction picture, which gives
 \begin{eqnarray} 
\langle \Phi(t) | \hat{\psi} (\mathbf{r}_1,t) | \Phi(t) \rangle &=& \langle \Phi_0 | \hat{U}^\dagger \hat{\psi} (\mathbf{r}_1,t) \hat{U} | \Phi_0 \rangle 
 .  
    \label{p1prime38}
\end{eqnarray}
As shown in the text, the second-order contribution to Eq.\ \eqref{p1prime38} can be rewritten as 
  \begin{eqnarray}
 \langle \hat{\psi} (\mathbf{r}_1,t) \rangle _I  &=&
     \langle  \Phi_0| \hat{U}^{\dagger(0)} \, [ \hat{\psi} (\mathbf{r}_1,t), \hat{{U}}^{(2)}  ] | \Phi_{0} \rangle \nonumber \\[2pt]      
     &+&\langle  \Phi_0| \hat{U}^{\dagger(1)} \,[ \hat{\psi} (\mathbf{r}_1,t) , \hat{{U} ^{(1)}}  ] | \Phi_{0}
     \rangle \nonumber \\[2pt] 
     &+&\langle  \Phi_0 | \hat{U}^{\dagger(2)} \,[ \hat{\psi} (\mathbf{r}_1,t), \hat{{U} ^{(0)}}  ] | \Phi_{0} \rangle   .   \label{threeterms}
 \end{eqnarray}
 Here $\hat{U} ^{(0)}$, $\hat{U} ^{(1)}$, and $\hat{U} ^{(2)}$ denote the zero, first, and second-order terms in an expansion analogous to Eq.\ \eqref{expansion}, with a similar notation for $\hat {U} ^\dagger$.

 The last term in Eq.\ \eqref{threeterms} is zero since $\hat{\psi} (\mathbf{r}_1,t)$ commutes with $\hat{{U} ^{(0)}}=\hat{I}$.  The second term is given by
 \begin{equation}
 \hspace{-0.1cm}
\begin{aligned}
 &  \langle \hat{\psi} (\mathbf{r}_1,t) \rangle    =  \langle  \,  \int d^3 \mathbf{r}' \,\int_{t_i}^{t} dt'    \hat{\mathbf{A}} (\mathbf{r}',t') \cdot \hat{\mathbf{j}} (\mathbf{r}',t')  \\[2pt]
  & \times   [\hat{\psi} (\mathbf{r}_1,t),  \int d^3 \mathbf{r}'' \,\int_{t_i}^{t} dt''    \hat{\mathbf{A}} (\mathbf{r}'',t'') \cdot \hat{\mathbf{j}} (\mathbf{r}'',t'') \,]  \rangle. 
     \label{secondterm}
\end{aligned}
 \end{equation}
 It can be shown that Eq.\ \eqref{secondterm} is proportional to the Hadamard function $\Delta_1^{\mu \nu}(x-x')\equiv \langle 0 |\{ \hat{A}^\mu(x),\hat{A}^\nu(x')  \} | 0 \rangle$ due to the fact that the integrals in Eq.\ \eqref{secondterm} are not time ordered.  As illustrated in Fig. B1, there are contributions from both $\hat{A} (\mathbf{r}_a,t_a) \hat{A} (\mathbf{r}_b,t_b) $  and  $\hat{A} (\mathbf{r}_b,t_b) \hat{A} (\mathbf{r}_a,t_a) $ for an arbitrary pair of spacetime points $\mathbf{r}_a,t_a$ and $\mathbf{r}_b,t_b$.  It can be shown that the coefficients of these two terms are equal, since the product of the two currents is the same in both cases and the commutator gives $\hat{\psi} (\mathbf{r}_1,t)$ as shown below .  This gives a total contribution proportional to $\Delta_1(x-x')$, which is real and does not contribute to the phase.  This result is analogous to the $\Delta_1$ term in Eq.\ \eqref{advancedretarded} that was neglected for the same reason.

\begin{figure}[btp]
\centering
\includegraphics[width=0.30\textwidth]{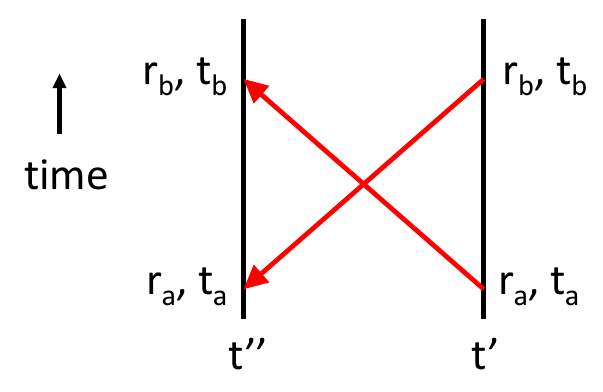}
\qquad
\caption{Pairs of contributions in the integrals over $t'$ and $t''$ in Eq.\ \eqref{secondterm}.  Here $\mathbf{r}_a,t_a$ and  $\mathbf{r}_b,t_b$  are two arbitrary spacetime points that can appear in both integrals.   The sum of these two contributions is proportional to $\Delta_1^{\mu \nu}(x-x')\equiv \langle 0 |\{ \hat{A}^\mu(x),\hat{A}^\nu(x')  \} | 0 \rangle$ , which is real and does not contribute to the phase shifts in the interferometer.  }
\label{Fig B1}
\end{figure}
 
  The remaining term in Eq.\ \eqref{threeterms} gives 
   \begin{eqnarray}
 \langle \hat{\psi} (\mathbf{r}_1,t) \rangle _I  &=&
     \langle  \Phi_0|  \, [ \hat{\psi} (\mathbf{r}_1,t), \hat{{U}}^{(2)}  ] | \Phi_{0} \rangle   ,   \label{commutator1}
 \end{eqnarray}
 as stated in the text.  Using the same techniques as in  Section III gives
 \begin{eqnarray}
     \hat{{U}}^{(2)} & = & \frac{i}{2}\  \int d^3 \mathbf{r}' \,\int_{t_i}^{t_f} dt' \int d^3 \mathbf{r}''  \int_{t_i}^{t'} dt''   \nonumber \\[2pt]
     & \times &  \hat{j}_l (\mathbf{r}',t')  \, G_R^{lm} (\mathbf{r}',t';  \mathbf{r}'',t'')   \hat{j}_m (\mathbf{r}'',t''), 
     \label{result78}
 \end{eqnarray}
 where $G_R^{lm} (\mathbf{r}',t';  \mathbf{r}'',t'')$ is the retarded propagator for the electromagnetic field and the expectation value in the initial (vacuum) state of the field has already been taken.  Combining Eqs.\ \eqref{commutator1} and \eqref{result78} gives
 \begin{equation}
 \begin{aligned}
    & \hspace{-0.2cm} \langle \hat{\psi} (\mathbf{r}_1,t) \rangle _I =  \frac{i}{2}\  \int d^3 \mathbf{r}' \,\int_{t_i}^{t_f} dt' \int d^3 \mathbf{r}''  \int_{t_i}^{t'} dt''    \\[2pt]
     & \times   \langle [\hat{\psi} (\mathbf{r}_1,t) ,\hat{j}_l (\mathbf{r}',t')  \, G_R^{lm} (\mathbf{r}',t';  \mathbf{r''},t'')   \hat{j}_m (\mathbf{r}'',t'') ] \rangle. 
     \end{aligned}
     \label{commutatorint}
 \end{equation} 
Using the identity $[\hat{A},\hat{B} \hat{C}]=[\hat{A},\hat{B}] \hat{C}+\hat{B}[\hat{A}, \hat{C}]$ gives the commutator of $\hat{\psi} (\mathbf{r}_1,t)$ with both $\hat{j}_l (\mathbf{r}',t')$ and $\hat{j}_m (\mathbf{r}'',t'')$.

First consider the contribution $ \langle\hat{\psi} (\mathbf{r}_1,t) \rangle _l$ from the commutator of $\hat{\psi} (\mathbf{r}_1,t)$ with  $\hat{j}_l (\mathbf{r}',t')$, which corresponds to the process shown in Fig. 2(a).  The commutator between the field at location $\mathbf{r}_1$  in the interferometer with the field at $\mathbf{r}_S$ in the source S is negligible for large separations, so that $\hat{j}_l (\mathbf{r}',t') $ can be identified with the current density $J_e$ in the interferometer.  Neglecting self-interaction terms, $\hat{j}_m (\mathbf{r}'',t'') $ must therefore correspond to the current density $J_S$ in the source. (More precisely, the integral over $\mathbf{r}'$ can be confined to a region $R_1$ that encloses path 1 and the integral over $\mathbf{r}''$ can be confined to a region $R_S$.) Using Eq.\ \eqref{result8} reduces Eq.\ \eqref{commutatorint} to 
 \begin{equation}
 \begin{aligned}
    & \hspace{-0.4cm} \langle \hat{\psi} (\mathbf{r}_1,t) \rangle _l =  \frac{i}{2}\  \int d^3 \mathbf{r}' \,\int_{t_i}^{t_f} dt'    \\[2pt]
     & \times   \langle [\hat{\psi} (\mathbf{r}_1,t) ,\hat{j}_l (\mathbf{r}',t')]    \hat{A}_{S}^l (\mathbf{r}',t')  \rangle. 
     \end{aligned}
     \label{commutatorint222}
 \end{equation} 
  
For an electron in path 1 described by a wave packet with a relatively well-defined velocity $\mathbf{v}_1(t')$, the current density is  given approximately by
$\hat{j} (\mathbf{r}',t') \approx q\mathbf{v}_1(t') \bar{\psi}  (\mathbf{r}',t' ) \hat{\psi}  (\mathbf{r}',t' )$.  In that case, Eq.\ \eqref{commutatorint222} becomes
   \begin{eqnarray}
\langle \hat{\psi} (\mathbf{r}_1,t) \rangle _l   &=& \frac{iq}{2}
      \int d^3 \mathbf{r}'   dt' \langle \,  [ \hat{\psi} (\mathbf{r}_1,t),  \bar{\psi} (\mathbf{r}',t') \hat{\psi} (\mathbf{r}',t') ] \,\rangle
         \nonumber \\[2pt]
     &\times& 
     \mathbf{v}_1(t') \cdot \mathbf{\mathcal{A}}_S (\mathbf{r}',t')    .  
 \label{commutator3}
 \end{eqnarray}

 It is straightforward to show that
 \begin{eqnarray}
     &&[ \hat{\psi} (\mathbf{r}_1,t),  \bar{\psi} (\mathbf{r}',t') \hat{\psi} (\mathbf{r}',t') ]
         \nonumber \\[2pt]
         &&=\{ \hat{\psi} (\mathbf{r}_1,t),  \bar{\psi} (\mathbf{r}',t')\} \hat{\psi} (\mathbf{r}',t'). 
         \label{anticommutator}
 \end{eqnarray} 
 This can be related to the retarded propagator $S_R(\mathbf{r}_1,t;\mathbf{r}',t')$ for the free Dirac field, which is defined \cite{peskin1995} by
 \begin{eqnarray}
     S_R(\mathbf{r}_1,t;\mathbf{r}',t')=  \Theta(t -t') \{ \hat{\psi} (\mathbf{r}_1,t),  \bar{\psi} (\mathbf{r}',t')\} 
         \label{Diracpropagator}.
 \end{eqnarray} 
 The unit step function $\Theta(t -t')=1$ inside the integral in Eq.\ \eqref{commutator3}, which allows it to be rewritten as
    \begin{eqnarray}
 \langle \hat{\psi} (\mathbf{r}_1,t) \rangle _l   &=& \frac{iq}{2}
      \langle \ \int d^3 \mathbf{r}'   dt'   \mathbf{v}_1(t') \cdot \mathcal{A}_S (\mathbf{r}',t') 
         \nonumber \\[2pt]
     &\times&   S_R(\mathbf{r}_1,t;\mathbf{r}',t') 
     \gamma^0 \hat{\psi} (\mathbf{r}',t')  \,\rangle
        .  
 \label{commutator4}
 \end{eqnarray}
 Here it has been assumed that the electrons are nonrelativistic, in which case the Dirac matrix $\gamma^0 $  is equivalent to the identity operator and can be inserted into the equation.
 
 The  Dirac propagator can be used to obtain the free field at time $t$ from the field at time $t'$ \cite{itzykson1980}: 
 \begin{equation}
     \hat{\psi} (\mathbf{r}_1,t)=\int d^3 \mathbf{r}' S_R(\mathbf{r}_1,t;\mathbf{r}',t') \gamma^0 \hat{\psi} (\mathbf{r}',t').
     \label{freefieldprop}
 \end{equation}
 Inserting this into Eq.\ \eqref{commutator4} gives
   \begin{eqnarray}
   \hspace{-0.6cm}
\langle  \hat{\psi} (\mathbf{r}_1,t) \rangle _l     &=& \frac{iq}{2}
       \int dt'\boldsymbol{v}_1(t') \cdot \mathcal{A}_S (t')  \langle \hat{\psi}_0 (\mathbf{r}_1,t)    \rangle  
       .  
 \label{lsubi}
 \end{eqnarray}
 Here it has been assumed that the vector potential is slowly varying over the dimensions of the wave packet, in which case it can be replaced by its average $\mathcal{A}_S (t')$ and taken outside the integral over $\mathbf{r}'$.   
 
 Including the first term in Eq.\ \eqref{p1prime499t} gives
   \begin{eqnarray}
   \hspace{-0.9cm}
 &&\langle \hat{\psi} (\mathbf{r}_1,t) \rangle _l 
  \nonumber \\[2pt]
  && = 
       [ 1+ \frac{iq}{2} \int dt' \boldsymbol{v}_1(t') \cdot \mathcal{A}_S (t')  ] \langle \hat{\psi}_0 (\mathbf{r}_1,t)    \rangle  
    .  
 \label{lsubi23}
 \end{eqnarray}
  It is apparent from Eq.\ \eqref{lsubi23} that the effect of the commutator $[\hat{\psi} (\mathbf{r}_1,t) ,\hat{j}_l (\mathbf{r}',t')] $ is to multiply the field of the electron in path 1 by a phase shift that corresponds to the first term in Eq.\ \eqref{result12p}.

The contribution from the  commutator of $\hat{\psi} (\mathbf{r}_1,t)  $ with $\hat{j}_m (\mathbf{r}'',t'') $ in Eq.\ \eqref{commutatorint} can be found in a similar way.  Equation \eqref{commutatorint} becomes 
 \begin{equation}
 \begin{aligned}
    & \hspace{-0.4cm} \langle \hat{\psi} (\mathbf{r}_1,t) \rangle _m =  \frac{i}{2}\  \int d^3 \mathbf{r}' \,\int_{t_i}^{t_f} dt' \int d^3 \mathbf{r}''  \int_{t_i}^{t'} dt''    \\[2pt]
     & \times   \langle \hat{j}_l (\mathbf{r}',t')  \, G_R^{lm} (\mathbf{r}',t';  \mathbf{r}'',t'')  [\hat{\psi} (\mathbf{r}_1,t) , \hat{j}_m (\mathbf{r}'',t'') ] \rangle. 
     \end{aligned}
     \label{commutatorint2}
 \end{equation} 
 The commutator gives $S_R \hat{\psi} (\mathbf{r}'',t'')$ as before, and the integral over $d^3 \mathbf{r}''$ can be performed as in Eq.\ \eqref{freefieldprop} to give $\hat{\psi} (\mathbf{r}_1,t)$, provided that $G_R^{lm} (\mathbf{r}',t';  \mathbf{r}'',t'') $ is assumed to be nearly constant over the dimensions of a wave packet and taken outside of the integral.  Here $\hat{j}_l (\mathbf{r}',t')$ corresponds to $\mathbf{J}_S$ while $\hat{j}_m (\mathbf{r}'',t'')$ corresponds to $\mathbf{J}_e$.  The result is that
   \begin{eqnarray}
   \hspace{-0.6cm}
 \langle \hat{\psi} (\mathbf{r}_1,t) \rangle _m   &=& 
       [ 1+ \frac{iq}{2} \int d^3\mathbf{r}'  dt' \boldsymbol{J}_S(\mathbf{r}',t') \cdot \mathcal{A}_e (\mathbf{r}',t')  ]
        \nonumber \\[2pt]
       &\times& \langle \hat{\psi}_0 (\mathbf{r}_1,t)    \rangle  
    .  
 \label{lsubi2}
 \end{eqnarray}
 This corresponds to the phase shift from the second term in Eq.\ \eqref{result12p}.

 The output of the beam splitter in path $1'$ is assumed to be an equal superposition of the fields in paths $1$ and $2$ as in the text.  Eq.\ \eqref{p1prime492} then gives 
\begin{eqnarray}
 P_{1'}=  \cos^2[(\phi_1 -\phi_2)/2],
\label{Last44}
\end{eqnarray}
where $\phi_1$ and $\phi_2$ are the phase shifts along the two paths as given by Eq.\ \eqref{result12p}.  

Equations \eqref{commutator3} and \eqref{lsubi}  depend on the assumption that the electron in the interferometer is described by a localized wave packet.  The same assumption is also required for the reasons discussed in Appendix A. This is not a fundamental limitation on the results, since a localization assumption of some kind is necessary in order to consider causality in the interaction between two separated particles. 

These results show that  in-in perturbation theory gives the same results as were obtained in the text by extracting the phase shift directly from the time-evolution operator. Eq.\ \eqref{commutator1} also shows that it is the commutator of the field with the higher-order terms in $U(t,t')$ that determines the phase shift in the interferometer.  Those commutators can be nonzero even when the current in the source and the field in the interferometer are spacelike separated with  $\{ \hat{\psi} (\mathbf{r}_S,t'), \bar{\psi}(\mathbf{r}_1,t) \}=0$.

\bibliographystyle{apsrev4-2}   
\bibliography{references}

\end{document}